\documentclass[preprint]{aastex}

\newcommand{\be}{\begin{equation}}
\newcommand{\ee}{\end{equation}}

\begin{document}

\title{DEUTERIUM CHEMISTRY IN THE PRIMORDIAL GAS}

\author{Daniele Galli and Francesco Palla}

\affil{Osservatorio Astrofisico di Arcetri \\
              Largo Enrico Fermi 5 \\
              I-50125 Firenze, Italy \\
              galli,palla@arcetri.astro.it}

\begin{abstract}

We review and update some aspects of deuterium chemistry in the
post-recombination Universe with particular emphasis on the formation
and destruction of HD. We examine in detail the available
theoretical and experimental data for the leading reactions of
deuterium chemistry and we highlight the areas where improvements in
the determination of rate coefficients are necessary to reduce the
remaining uncertainties.  We discuss the cooling properties of HD
and the modifications to the standard cooling function
introduced by the presence of the cosmological radiation field.
Finally, we consider the effects of deuterium chemistry on the
dynamical collapse of primordial clouds in a simple ``top-hat''
scenario, and we speculate on the minimum mass a cloud must have in
order to be able to cool in a Hubble time.

\end{abstract} 


\section{Introduction}

The formation of H$_2$ and HD molecules in the post-recombination Universe
plays a central role in the evolution of gas condensations.  Even trace
abundances of these molecules strongly affect the cooling properties of
the primordial gas which would be otherwise an extremely poor radiator
(cooling by Ly-$\alpha$ photons is ineffective at temperatures
less than $\sim 10^4$~K).

The chemistry of the early Universe has been investigated in several
studies starting with the seminal paper by Lepp \& Shull~(1983). We
mention in particular the work of Puy et al.~(1993), Galli \&
Palla~(1998; hereafter GP) and the comprehensive analysis of deuterium
chemistry by Stancil, Lepp \& Dalgarno~(1998; hereafter SLD).  The
abundances of H$_2$ and HD predicted at low redshift in these studies
are of the order of $10^{-6}$ and $10^{-9}$, respectively, depending on
the cosmological model adopted (see Palla, Galli \& Silk~1995 for the
variation of the chemical abundances with the assumed baryon-to-photon
ratio).  A comparison of the abundance of HD obtained at $z=10$ by 
GP ($n[{\rm HD}]/n[{\rm H}]=1.1\times 10^{-9}$), and SLD ($n[{\rm
HD}]/n[{\rm H}]=1.6 \times 10^{-9}$) for the same cosmological model
($h=0.5$, $\Omega_0=1$, $\Omega_b=0.0367$) shows satisfactory
agreement, although this might not be completely significant since both
calculations were based on a compilation of reaction rates obtained
basically from the same sources. To repeat the words of SLD, {\it
``further studies are needed to reduce the uncertainty in the HD
abundance''}. One of the goals of this paper is to review the progress
made in our understanding of HD chemistry in the three years subsequent
to the publication of these studies. We will outline the improvements
which have occurred in the meantime, and discuss the remaining uncertainties.

From an observational standpoint, the abundance of atomic deuterium and
molecular hydrogen has been measured in several cosmological clouds at
redshift $z\simeq 2$--3 in absorption along the line of sight to bright
quasars.  Deuterium in particular has attracted renewed
attention because of the controversy about its abundance in
high-redshift Ly-$\alpha$ clouds (see Hogan~1998 for an account and a
resolution of the controversy).  Molecular hydrogen at high redshift
was first detected by Levshakov \& Varshalovich~(1985) in a damped
Ly-$\alpha$ system at $z=2.8$. Since then, the presence of H$_2$ has
been confirmed in at least four additional systems (see e.g. Levshakov
et al.~2000 and references therein).  

In Fig.~1 we compare observational data on D and H$_2$ at high redshift
with the corresponding abundances calculated with the standard model of
GP that follows the homogenous expansion of the universe. The agreement
between the theoretical and observed deuterium abundance is excellent.
In the case of H$_2$, one should keep in mind the sensitivity of this
molecule to ambient conditions in Ly-$\alpha$ clouds, such as the
stronger ultraviolet radiation field and the lower dust content with
respect to local interstellar medium. These fctors can account for the
considerable spread of observed abundances. Despite the complex
phenomenology associated with damped Ly-$\alpha$ systems and the
resulting uncertainty in the interpretation of the results, it is
encouraging, to say the least, to witness the emergence of the
observational foundations of the highly theoretical discipline of
primordial chemistry.

\section{Chemical Reactions}

The formation of HD in the primordial gas follows two main routes,
involving a deuteron exchange with H$_2$:
\be
{\rm D}+{\rm H}_2\rightarrow \rm{HD}+{\rm H},
\label{dh2}
\ee
and
\be
{\rm D}^+ +{\rm H}_2\rightarrow \rm{HD}+{\rm H}^+.
\label{dph2}
\ee

Being an isotopic modification of the the most fundamental
three-electron interaction, namely, the ${\rm H}+{\rm H}_2$ reaction,
reaction~(\ref{dh2}) has received considerable interest.  Thermal rate
constants for this reaction have been measured by Ridley, Schulz, \&
LeRoy~(1966), Westenberg \& de Haas~(1967), Mitchell \& LeRoy~(1973),
and Michael \& Fisher~(1990) over a wide range of temperatures.
Theoretical calculations employing statistical, semiclassical and
quantal method have been performed by several groups, and show very
good agreement with each other and with the experimental data. The most
recent studies are by Zhang \& Miller~(1989), Michael, Fisher, \&
Bowman~(1990), Mielke et al.~(1994), and Charutz, Last, \& Baer~(1997).
These results are compared in the usual Arrhenius plot shown in Fig.~2
(the high-temperature experimental results of Michael \& Fisher~1990
are not shown). In this paper we adopt the reaction rate computed by
Mielke et al.~(1994) with the DMBE surface (Varandas et al.~1987) which
predicts much more accurate low-temperature kinetics than other
surfaces. Their results agree with the values computed by Michael \&
Fisher~(1990) (with the same surface) within $\sim 20\%$.  We also note
that Zhang \& Miller~(1989) computed state-by-state cross sections for
$vJ\rightarrow v^\prime J^\prime$ transitions and state-to-state rate
constants in the temperature range 200--1000~K. The agreement with the
experimental cross section values at $E\simeq 1$~eV (Phillips, Levene,
\& Valentini~1989) is within a factor $\sim 2$.

Reaction~(\ref{dph2}) represents the major source of HD in diffuse
interstellar clouds (Dalgarno, Weisheit, \& Black~1973). Its rate
coefficient is almost constant and close to the Langevin value ($2\times
10^{-9}$~cm$^3$~s$^{-1}$, see Fig.~3).  The reaction rate has been
measured in the laboratory by Fehsenfeld et al.~(1973, at $T=200$ and
278~K) using a flowing afterglow technique, and by Henchman, Adams,
\& Smith~(1981, at $T=205$ and 295~K) using a variable-temperature
selected ion flow tube technique (see also Smith, Adams, \&
Alge~1982).  Gerlich~(1982) performed accurate quantum-mechanical
calculations of the rate coefficient, which are in excellent agreement
with the experimental results of Henchmann et al.~(1981) at $T=295$~K, but
less at $T=205$~K (still within a factor $\sim 2$). There is a discrepancy
with the results of Fehsenfeld et al.~(1973) at both temperatures.

The destruction of HD occurs mainly via the reverse reactions of 
(\ref{dh2}) and (\ref{dph2}),
\be
\rm{HD}+{\rm H} \rightarrow {\rm D}+{\rm H}_2,
\label{hdh}
\ee
and 
\be
\rm{HD}+{\rm H}^+ \rightarrow {\rm D}^+ +{\rm H}_2.
\label{hdhp}
\ee
In general, the rate coefficients for the forwards and reverse 
chemical reactions are related by the standard thermodynamic expression
(e.g. Berry, Rice \& Ross~1980)
\be
\ln\left(\frac{k_{\rm f}}{k_{\rm r}}\right)=-\frac{\Delta H^0}{RT}+\frac{\Delta S^0}{R},
\label{detbal}
\ee
where $R$ is the universal gas constant, and $\Delta H^0$, $\Delta S^0$
are the enthalpy and entropy changes. Thus, if the rate $k_{\rm f}$ is
knwon, the rate for the reverse reaction $k_{\rm r}$ can be obtained
directly from eq.~(\ref{detbal}). 
From the differences in the zero point vibrational
energies of H$_2$ and HD and in the ionization potentials of H and D,
one obtains the enthalpy changes for reactions~(\ref{dh2}) and (\ref{dph2}): 
$\Delta H^0/R=-412$~K and $-462$~K, respectively.
The entropy change $\Delta S^0/R$ can be calculated on statistical
grounds (see e.g. Flower~2000) and is the same for both reactions: $\Delta
S^0/R=\ln 2$ for para-H$_2$, $\Delta S^0/R=\ln (2/3)$ for ortho-H$_2$.
These values are for reactant and product molecules in their ground states.
Both the entropy and the enthalpy changes are modified when
rovibrationally excited molecular states are involved (Flower~2000).

Rather than using eq.~(\ref{detbal}), in this paper we prefer to
provide independent fitting formulae for both the direct and reverse
reactions that dominate the chemistry of deuterium in the primordial
gas. For specific applications, the reader may adopt the reaction rate
of the forward (or reverse, if better constrained) reaction, and
compute the rate of the reverse (forward) reaction from the
thermodynamic relation expressed by eq.~(\ref{detbal}).

The rate coefficient of reaction (\ref{hdh}) has been calculated by
Shavitt~(1959) using a semiempirical H$_3$ energy surface. Only sparse
laboratory data exist for this reaction. At $T=10^3$~K, the
experimental result of Boato et al.~(1956) is within $\sim 20\%$ from
the value calculated by Shavitt~(1959).  At lower temperatures
(720--880~K), the theoretical rate is a factor $\sim 2$ lower than the
experimental data by van Meersche~(1951), but the extrapolation of the
adopted H$_3$ energy surface introduces significant uncertainty in the
results at low temperatures (see Fig.~4). As a challenge to chemical
physicists, we recall the words of Shavitt~(1959): {\it ``it is
unfortunate that for a reaction as basically important as the one
considered here, the experimental data are so incomplete and
inconclusive''}.

Since reaction~(\ref{hdhp}) is endothermic by 0.0398~eV (462~K), the
removal of HD at low temperatures is reduced by a factor
$\exp(-462/T)$, and this can lead to significant enhancement of the
HD/H$_2$ ratio (fractionation). The rate coefficient of this reaction
has been measured in the laboratory by Henchmann et al.~(1981) at
$T=205$ and 295~K, and the results are in good agreement with the values
obtained by applying the principle of detailed balance to the reverse
reaction (\ref{dph2}). As in GP, we adopt the rate coefficient
calculated by Gerlich~(1982) over the temperature range 30--600~K (see
Fig.~5).

Finally, the relative abundance of D and D$^+$ is determined by the charge exchange 
reactions 
\be
{\rm H}^++{\rm D} \rightarrow {\rm H}+{\rm D}^+
\label{hpd}
\ee
and its reverse
\be
{\rm H}+{\rm D}^+ \rightarrow {\rm H}^++{\rm D}.
\label{hdp}
\ee
The cross section for reaction~(\ref{hpd}) was computed by
Matveenko~(1974) and Hunter \& Kuriyan~(1977) for energies from $10^{-3}$
to 7.5~eV. The two results differ by a factor $\sim 2$ at low energies
for reasons unclear. Subsequent calculations by Hodges \& Breig~(1993),
and, more recently, by Igarashi \& Lin~(1999) and Zhao, Igarashi, \& Lin~(2000),
confirm the validity of the results of Hunter \& Kuriyan~(1977),
and improve significantly the accuracy of the calculations around
$\sim 10^{-3}$~eV. Good agreement was also found with the experimental
measurements of Newman et al.~(1982) and Esry et al.~(2000). The cross
section of reaction~(\ref{hdp}) was also calculated by Igarashi \&
Lin~(1999) and Zhao et al.~(2000).  Unfortunately, no experimental data
are available for this reaction.

Watson~(1976) estimated the rate of reactions (\ref{hpd}) and
(\ref{hdp}) and Watson, Christensen \& Deissler~(1978) calculated
the rate coefficient on the basis of the cross section obtained by Hunter \&
Kuriyan~(1977). These rates have been widely adopted in studies
of deuterium chemistry.  Since reaction~(\ref{hpd}) has a threshold
of $43$~K, the rate coefficient for reaction (\ref{hdp}) is usually
obtained by multiplying that of reaction~(\ref{hpd}) by $\exp(43/T)$.
The situation has been reanalyzed recently by Wolf Savin (2001) who
computed accurate rates for reactions~(\ref{hpd}) and (\ref{hdp}) from
the cross sections of Igarashi \& Lin~(1999) and Zhao et al.~(2000). These
results are compared to those of Watson et al.~(1978) in Figs.~6 and 7.

In Table~1 we collect accurate fitting formulae (computed by us or by
the authors quoted in the references) for the chemical reactions
discussed in this section. These expressions update
and replace the corresponding formulae given in Table~2 of GP.

In addition to these reactions, additional contributions to the formation of HD
in the early Universe come from the associative detachment reactions
\be
{\rm D}+{\rm H}^-\rightarrow {\rm HD}+e
\ee
and 
\be
{\rm D}^-+{\rm H}\rightarrow {\rm HD}+e,
\ee
whose rate coefficients however are not known, and can only be
estimated from the corresponding H reactions (see SLD). Finally, 
minor contributions to the formation of HD come from the radiative
association reaction
\be
{\rm H}+{\rm D}\rightarrow {\rm HD} + h\nu,
\ee
whose rate coefficient was computed by Stancil \& Dalgarno~(1997),
and from reactions involving less abundant deuterated species like 
HD$^+$ and H$_2$D$^+$ (see SLD for details). 

Generally, the chemistry of HD in the primordial gas is dominated by
the ion--neutral reactions (\ref{dph2}) and (\ref{hdhp}) in a gas of
low density (e.g. the primordial gas before the formation of the first
structures), whereas the neutral--neutral reactions (\ref{dh2}) and
(\ref{hdh}) become more important in conditions of high density and
temperature (cloud collapse, shocked gas).

\section{Heating and Cooling}

In order to calculate the cooling properties of HD, one should know the 
population of all rovibrational levels. In steady-state, these are obtained 
by solving the balance equations 
\be
x_J \sum_{J^\prime} [R_{JJ^\prime}(T_{\rm rad}) 
+ C_{JJ^\prime}(n,T_{\rm gas})] =\sum_{J^\prime}x_{J^\prime}[R_{J^\prime J}(T_{\rm rad}) +
C_{J^\prime J}(n,T_{\rm gas})],
\ee
where $J$ and $J^\prime$ indicate a generic couple of rovibrational
levels.  The collisional transition probabilities
$C_{JJ^\prime}(n,T_{\rm gas})$ and $C_{J^\prime J}(n,T_{\rm gas})$ are
obtained by multiplying the corresponding excitation coefficients, 
$\gamma_{JJ^\prime}(T_{\rm gas})$ and $\gamma_{J^\prime J}(T_{\rm
gas})$, and the density of the colliding species. The terms
$R_{JJ^\prime}$ and $R_{J^\prime J}$ are the radiative excitation and
de-excitation rates, that can be expressed in terms of the Einstein
coefficients $A_{JJ^\prime}$ and $B_{JJ^\prime}$,
\be
R_{JJ^\prime}=\cases
{A_{JJ^\prime}+B_{JJ^\prime}u(\nu_{JJ^\prime},T_{\rm rad}), & $J^\prime<J$,  \cr
B_{JJ^\prime}u(\nu_{JJ^\prime},T_{\rm rad}),                & $J^\prime>J$,  \cr}
\ee
where
$u(\nu_{JJ^\prime},T_{\rm rad})$ is the energy density of the cosmic background 
radiation (CBR) per unit frequency at the temperature $T_{\rm rad}$,
\be
u(\nu_{JJ^\prime},T_{\rm rad}) = \frac{8\pi h \nu_{JJ^\prime}^3}{c^2}
\left[\exp(h\nu_{JJ^\prime}/k T_{\rm rad})-1\right]^{-1}.
\ee

In the primordial gas, collisional excitation of HD is dominated by
collisions with H, and, to a less extent, with He.  The 
coefficients for inelastic scattering of He--HD were computed by
Green~(1974) and Schaefer~(1990) at temperatures $T\le 600$~K, for
$0\le J\le 3$ and $\Delta J= +1, +2$. Collisional coefficients for rotational
excitation of the system H--HD were usually derived by scaling the
He-HD values with the square root of the ratio of the reduced masses of
the two systems, $\gamma_{\rm H-HD}=(\mu_{\rm He-HD}/\mu_{\rm
H-HD})^{1/2}\gamma_{\rm He-HD}$, where $(\mu_{\rm He-HD}/\mu_{\rm
H-HD})^{1/2}=1.51$ (see e.g.  Timmerman~1996, GP).  In the words of
Timmermann~(1996), {\it ``this assumption is however an educated guess,
and data on H--HD are urgently needed''}.  These data were recently provided
by Flower \& Roueff~(1999) (for H and H$_2$) and Roueff \&
Zeippen~(1999) (for He), using full
quantum-mechanical methods and updated energy potential surfaces to
evaluate rovibrational excitation coefficients for collisions of HD
with H, H$_2$ and He. The results for H--HD collisions computed by
Flower \& Roueff~(1999) in the temperature range $100\le T_{\rm gas}
\le 2000$~K for $v\le 2$ and $J\le 9$ for a total of 30 rovibrational
levels are available from the CCP7 server {\tt
http:$/\!/$ccp7.dur.ac.uk$/$}.

The energy levels and the Einstein coefficients $A_{JJ^\prime}$ of HD
were calculated by Abgrall et al.~(1982) who considered both dipole and
quadrupole transitions and included a large number of rovibrational
levels.  Since the energy spacing of the rotational levels of HD
is quite large, $E_1/k=128$~K, $E_2/k=383$~K, $E_3/k=764$~K, etc.,
GP computed the HD cooling function with a simple four-level
system ($J=0$--3) adopting the collisional coefficients of
Schaefer~(1990).  Flower et al.~(2000) updated the calculations of GP
adopting the collisional rate coefficients of Flower \& Roueff~(1999)
and Roueff \& Zeippen~(1999). The HD cooling function computed by
Flower et al.~(2000) is also available from the CCP7 server.
A useful approximation in the low-density limit is the expression
\be 
\Lambda_{\rm HD}[n({\rm H}\rightarrow 0)]=2\gamma_{10}(T_{\rm gas})E_{10} e^{-E_{10}/kT_{\rm gas}}
+(5/3)\gamma_{21}(T_{\rm gas})E_{21} e^{-E_{21}/kT_{\rm gas}}, 
\ee 
where $E_{10}/k=128$~K, $E_{21}/k=255$~K and the collisional rate coefficients
$\gamma_{JJ^\prime}$ are given by Flower \& Roueff~(1999):
\be
\gamma_{10}(T_{\rm gas})=4.4\times 10^{-12}+3.6\times 10^{-13}(\log T_{\rm gas})^{0.77}~{\rm cm}^3~{\rm s}^{-1},
\ee
and
\be
\gamma_{21}(T_{\rm gas})=4.1\times 10^{-12}+2.0\times 10^{-13}(\log T_{\rm gas})^{0.92}~{\rm cm}^3~{\rm s}^{-1}.
\ee

The comparison between the HD cooling rate calculated by GP and Flower
et al.~(2000), shown in Fig.~8 is instructive. A simple four-level
molecule is able to predict quite accurately the cooling rate in a wide
range of temperatures ($T_{\rm gas} \lesssim 2000$~K) and densities
($n[{\rm H}]\lesssim 10^7$~cm$^{-3}$), but of course fails badly in the
high-temperature, high-density regime.  However, for cosmological
applications, it is important to keep in mind that the HD cooling
function of Flower et al.~(2000) has been computed assuming that the
temperature of the CBR is much smaller than the gas temperature.  This
approximation is valid at low redshifts (where the two temperatures
differ by more than a factor $\sim 10$ for $z\lesssim 20$), but becomes
increasingly inaccurate at higher redshifts.  Since most cosmological
scenarios of structure formation begin at $z\simeq 100$, when $T_{\rm
rad}\simeq 300$~K, the level population of molecules is strongly
affected by stimulated emission and absorption.  In such conditions,
molecules may become an effective heating source for the gas, because
the rate of collisional de-excitation of the rovibrational levels is
faster than their radiative decay (Khersonskii~1986; Puy et al.~1993).

As an illustration of this effect, we plot in Fig.~9 the net heat
transfer function $(\Gamma-\Lambda)_{\rm HD}$ computed with the GP
model for $n({\rm H})=1$~cm$^{-3}$ as function of gas temperature at
three selected redshifts. Note that for $T_{\rm gas}> T_{\rm rad}$ the
cooling function is significantly decreased from the value computed
with $T_{\rm rad}=0$, because of the radiative depopulation of excited
states. The sudden drop signals the condition $T_{\rm gas}=T_{\rm
rad}$.  Finally, for $T_{\rm gas}<T_{\rm rad}$ the function changes
sign and becomes a net heating term for the gas. In cosmological
simulations the heat transfer should be computed self-consistently at
each redshift.

\section{Application: cloud collapse at $z\simeq 100$}

In both cold dark matter and baryonic dark matter cosmological
scenarios, the first objects predicted to enter in the nonlinear stage
are the smallest ones. In cold dark matter models, it is expected that
overdense regions with masses $10^5$--$10^7$ $M_\odot$ first
collapse in the redshift range $10\la z \la 100$ (see e.g. Cen,
Ostriker, \& Peebles~1993). The crucial question is whether molecular
cooling allows the baryonic component to dissipate its kinetic energy
and collapse on a timescale shorter than a Hubble time.  This important
question is fully addressed e.g. in the 1-D hydrodynamical calculations
by Haiman, Thoul, \& Loeb~(1996) and in the 3-D numerical simulations
presented by Abel, Anninos, \& Norman~(1997) and Bromm, Coppi, \&
Larson~(1999).  Here, we consider a particular issue related to the
choice of the initial conditions for collapse calculations.

Following Tegmark et al.~(1997), we consider the growth of a ``top-hat''
overdensity region, an isolated spherical perturbation in a uniform
density Universe (see e.g. Padmanabhan~1993). The radius of this region
increases at a slower rate than the scale factor of the Universe (but
still obeys the Friedman's equations) and, after reaching a maximum
value (turnaround), recollapses to a point. The formation of a
singularity is clearly an artifact of the simplified assumptions of the
``top-hat'' model.  In a realistic situation, gas dynamical process
(internal pressure, shocks) will eventually halt the collapse of the
baryonic component at some finite value of the density. The resulting
quasi-equilibrium configuration is a virialized ``halo'', and its
subsequent evolution depends on the ability of the gas to cool on a
timescale shorter than a dynamical timescale.

We show in Fig.~10 the results of a sample run of the Tegmark et
al.~(1997) evolutionary model, obtained with our chemical network and
molecular cooling prescriptions. For the case shown, the cloud
reaches virialization at $z_{\rm vir}=110$, where the temperature is
$T_{\rm vir}=2000$~K. After virialization, the gas density is assumed
to remain constant and uniform for the rest of the run, and we follow
the cooling of the gas due to H$_2$ and HD molecules.  This is clearly
a poor approximation since the cooling of the gas will of course induce
an increase in the the density, and the post-virialization evolution of
the cloud must be followed with a full hydrodynamical calculation
(Galli et al., in preparation). Nevertheless, it is
instructive to consider how rapidly the baryons are able to dissipate
the thermal energy of the cloud via molecular cooling. As we see in
Fig.~10, recombination reduces the ionization fraction of the cloud to
negligible values, weakening the Compton cooling of the gas (since
$T_{\rm rad}< T_{\rm gas}$). The rapid rise of H$_2$ (via the H$^-$
channel) around $z_{\rm vir}$ causes the sudden cooling of the gas from
$T_{\rm vir}$ down to a few hundred degrees, and the subsequent
increase in the HD abundance further reduces the gas temperature down
to few tens of degrees, establishing again thermal coupling of gas and
cosmic radiation at redshift $z\simeq 10$.

Loosely speaking, a Hubble time corresponds to the redshift dropping by
a factor $2^{2/3}\simeq 1.6$. For the particular case considered in
this illustrative example, at a resdhift $z=z_{\rm vir}/1.6\simeq 70$
the cloud temperature has dropped to $T_{\rm gas} \simeq 350$, i.e.  of
a factor $\sim 6$ with respect to the virial temperature. Molecular
cooling (mostly H$_2$ in this range of redshift) therefore enables the
cloud to cool significantly within a Hubble time, and eventually to
collapse and form luminous objects. Cooling by HD appears to induce a
further substantial reduction of the temperature at later times, but
the subsequent evolution of the cloud can only be followed with a
realistic calculation. However, it must be kept in mind that the mass
scale of the fragments formed by the collapse of the cloud depends
sensitively on the gas temperature ($M_J\sim \rho^{-1/2}T_{\rm
gas}^{3/2}$). The cooling induced by the formation of HD may in fact
reduce the gas temperature below $\sim 100$~K and therefore enable the
formation of primordial low-mass stars or even brown dwarfs, a
possibility recently explored by Uheara \& Inutsuka~(2000) (see also
Flower \& Pineau des For\^ets~2001). These exciting results represent a
radical cange of perspective in the field of primordial star formation
and deserve further investigation.

\section{Conclusions}

We have examined the most effective chemical reactions leading to the
formation/destruction of HD molecules in the early Universe, and
presented a list of accurate reaction rate coefficients for primordial
chemistry calculations that update those adopted in previous studies.
We have analyzed the heating/cooling properties of HD molecules,
stressing the relevance of a self-consistent calculation of the energy
transfer rate between gas and radiation for cosmological applications.
As an illustration, we have presented a simplified model for
the evolution of density perturbations in the expanding Universe,
following previous investigations by Haiman et al.~(1996) and Tegmark
et al.~(1997), but including our comprehensive and updated chemical
network.  These results, together with recent findings by
Uheara \& Inutsuka~(2000) and Flower \& Pineau des For\^ets~(2001) 
underline the substantial contribution of HD to gas
cooling during the collapse of primordial clouds. A preliminary conclusion
from these studies is that HD is at least as important 
as H$_2$ in determining the thermal balance of zero-metal clouds.

\acknowledgements 

This work is supported by the Italian Ministry for the University and for
Scientific and Technological Research (MURST) through a COFIN-2000
grant. It is a pleasure to thank Guillaume Pineau des For\^ets 
and Malcolm Walmsley for informative dicussions on deuterium chemistry.
We also thank A. Dalgarno, S. N. Shore and an anonymous referee 
for useful comments on an earlier version of this paper.

\clearpage

\begin{deluxetable}{llll}
\tablewidth{0pt}
\tablecaption{\sc Rate coefficients}
\tablehead{
\colhead{\#} & \colhead{reaction} & \colhead{rate coefficient (cm$^3$~s$^{-1}$)} & \colhead{reference} \\
}
\startdata
(1) & ${\rm D}+{\rm H}_2\rightarrow {\rm HD}+{\rm H}$       & 
$1.69\times 10^{-10}e^{(-4680/T+198800/T^2)}$ 
						      & Mielke et al.~(1994) \\
(2) & ${\rm D}^+ +{\rm H}_2\rightarrow {\rm HD}+{\rm H}^+$  &  
$10^{-9}\times [0.417+0.846\log T-0.137 (\log T)^2]$
                                                      & Gerlich (1982)  \\
(3) & ${\rm HD}+{\rm H} \rightarrow {\rm D}+{\rm H}_2$      &
$5.25\times 10^{-11}e^{(-4430/T+173900/T^2)}$
                                                      & Shavitt (1959)  \\
(4) & ${\rm HD}+{\rm H}^+ \rightarrow {\rm D}^+ +{\rm H}_2$ & 
$1.1\times 10^{-9}e^{-488/T}$
						      & Gerlich (1982)  \\
(5) & ${\rm H}^++{\rm D} \rightarrow {\rm H}+{\rm D}^+$     & 
$2.00\times 10^{-10}T^{0.402}e^{-37.1/T}-3.31\times 10^{-17}T^{1.48}$
                                                      & Wolf Savin~(2001)  \\
(6) & ${\rm H}+{\rm D}^+ \rightarrow {\rm H}^++{\rm D}$     &
$2.06\times 10^{-10}T^{0.396}e^{-33.0/T}-2.03\times 10^{-9}T^{-0.332}$
                                                      & Wolf Savin~(2001)  \\
\enddata
\end{deluxetable}

\clearpage
\begin{figure}
\plotone{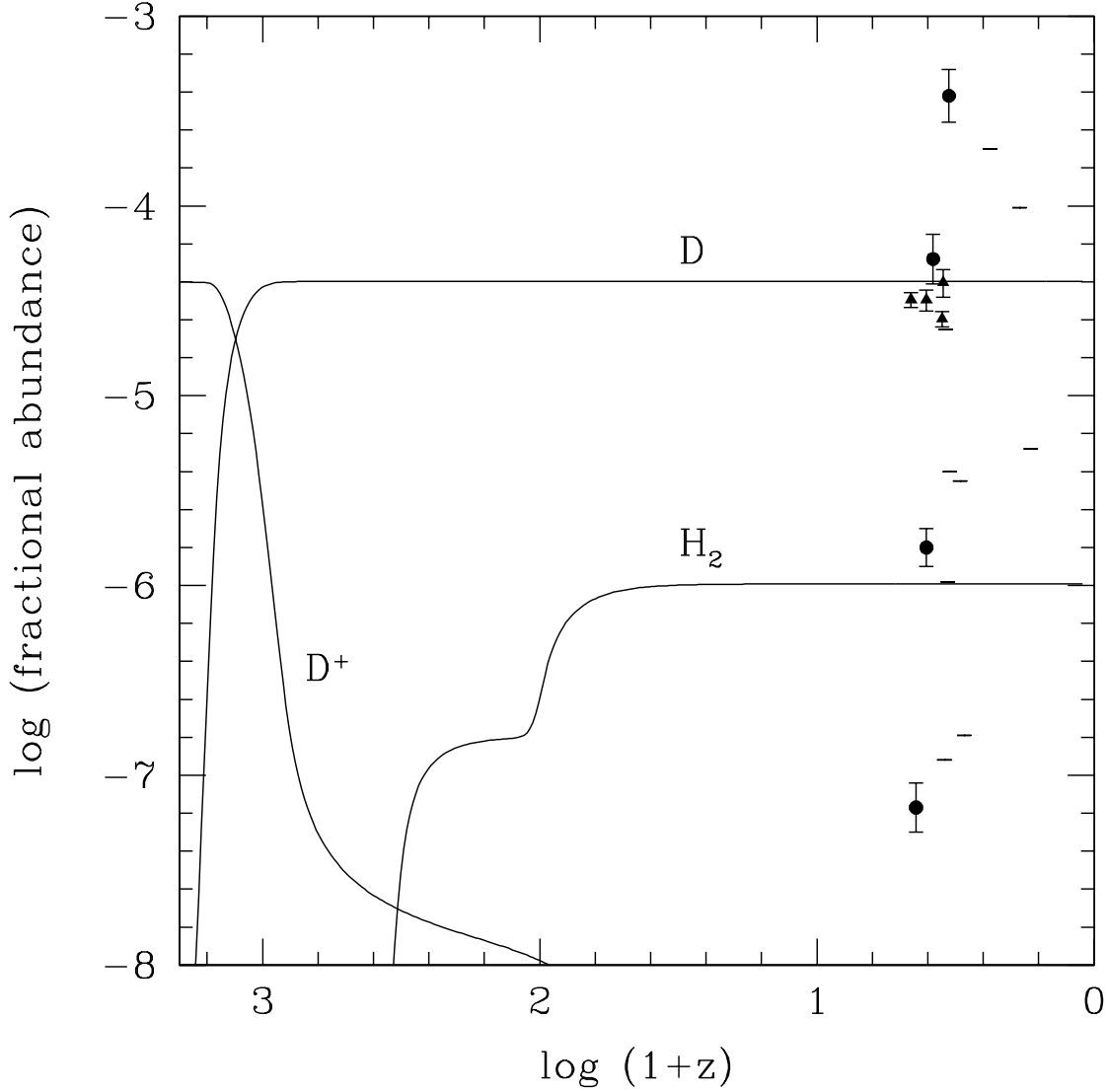} 
\caption{Comparison of the abundances
of D, D$^+$ and H$_2$ in the primordial gas, relative to hydrogen,
as function of redshift $z$
calculated with the standard model of GP.  Data points represent
abundance measurements of H$_2$ and D in damped Ly-$\alpha$ systems.
The H$_2$ data are taken from the list compiled by Levshakov et
al.~(2000) (detections, {\it dots with errorbars}; {\it dashes}: upper
limits), whereas D data ({\it triangles with errorbars}) are from
Burles \& Tytler~(1998a,b), O'Meara et al.~(2001) and D'Odorico et
al.~(2001).}
\end{figure}
\clearpage

\begin{figure}
\plotone{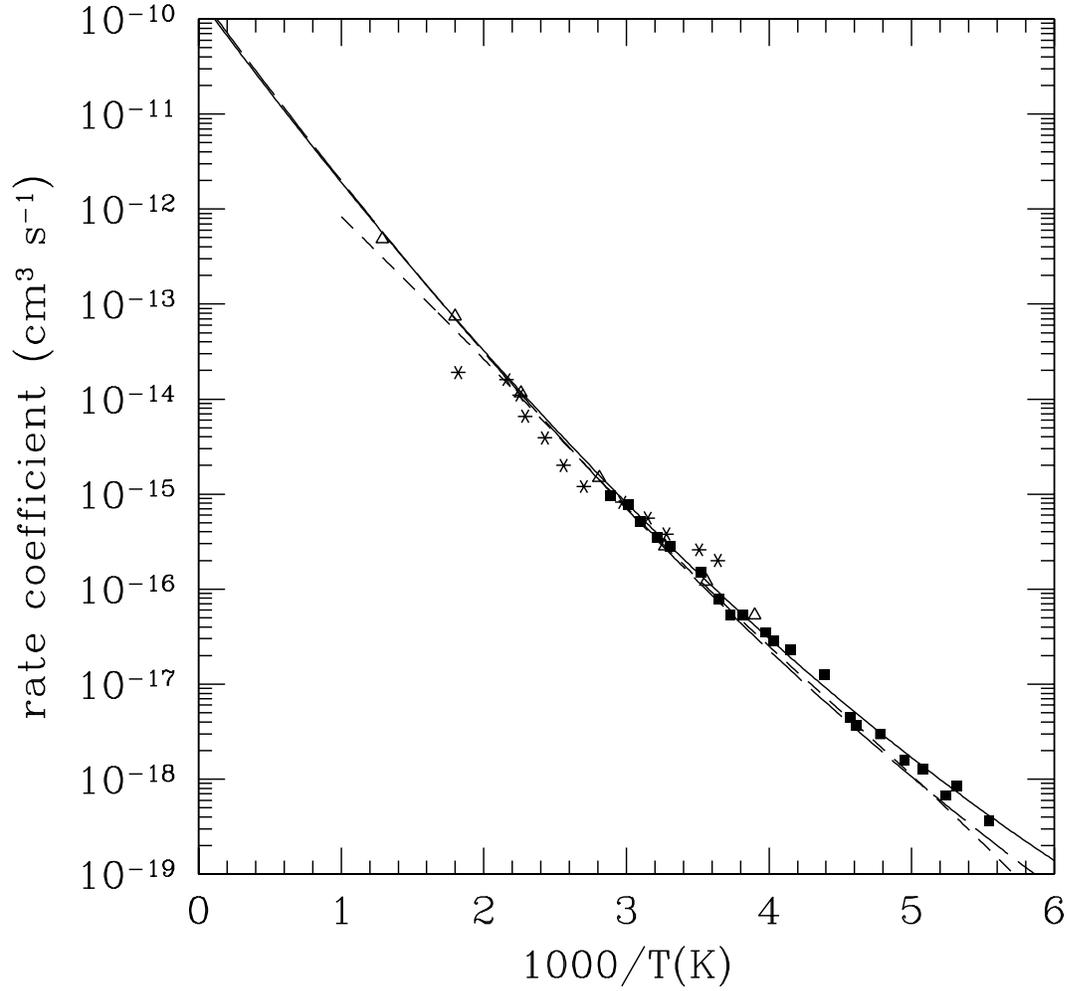}
\caption{Rate coefficient for the reaction ${\rm D}+{\rm
H}_2\rightarrow \rm{HD}+{\rm H}$ according the calculations of Charutz
et al.~(1997) ({\it short-dashed line}), 
Zhang \& Miller~(1989) ({\it long-dashed line}),
Mielke et al.~(1994) ({\it solid line}). 
Experimental data are from Mitchell \& LeRoy (1973) ({\it filled
squares}), Westenberg \& deHaas~(1967) ({\it open triangles}), Ridley
et al.~(1966) ({\it asterisks}).}
\end{figure}
\clearpage

\begin{figure}
\plotone{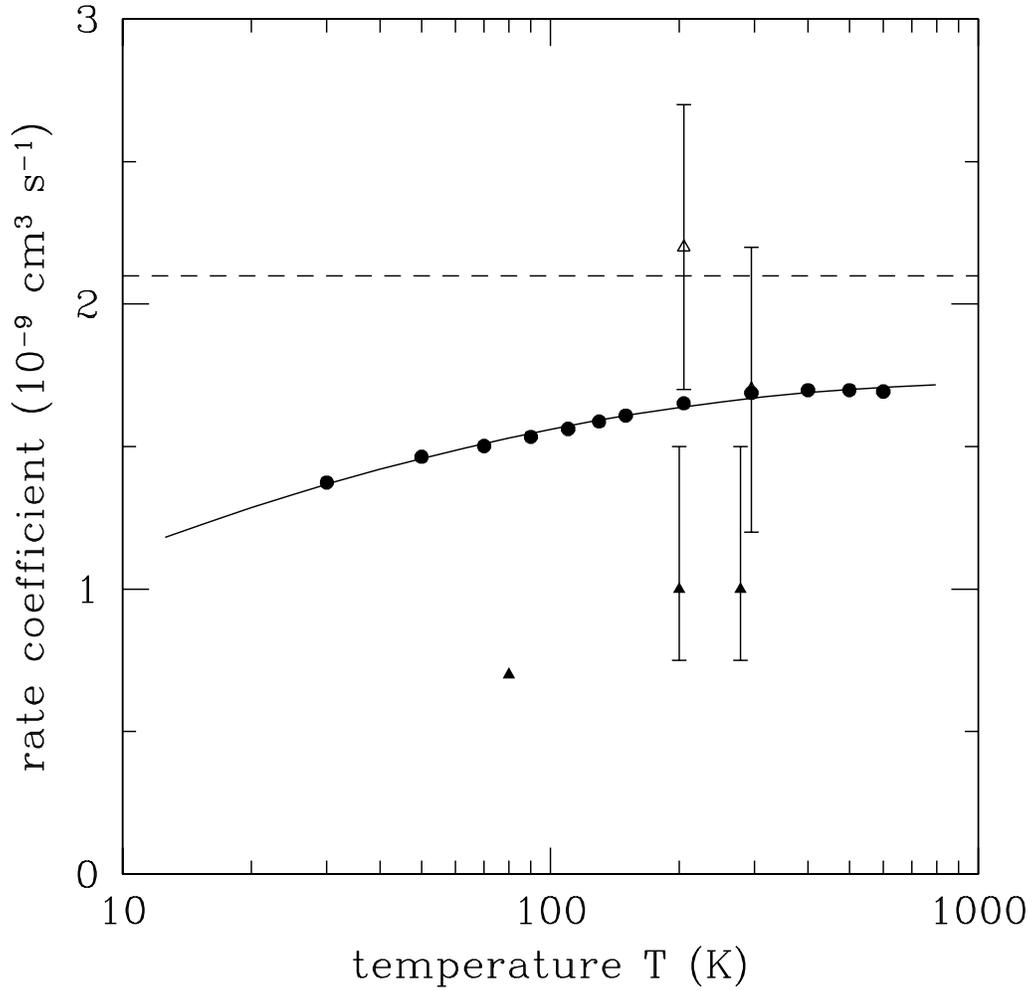}
\caption{Rate coefficient for the reaction ${\rm D}^+ +{\rm
H}_2\rightarrow \rm{HD}+{\rm H}^+$ according to the calculations of
Gerlich~(1982) ({\it filled dots}) and the measuruments by Fehsenfeld
et al.~(1973) ({\it filled triangles}) and Henchman et al.~(1981) ({\it
empty triangles}). The {\it dashed line} shows the Langevin value of
the rate coefficient. The {\it solid line} shows our fit to Gerlich's
results (see Table~1).}
\end{figure}
\clearpage

\begin{figure}
\plotone{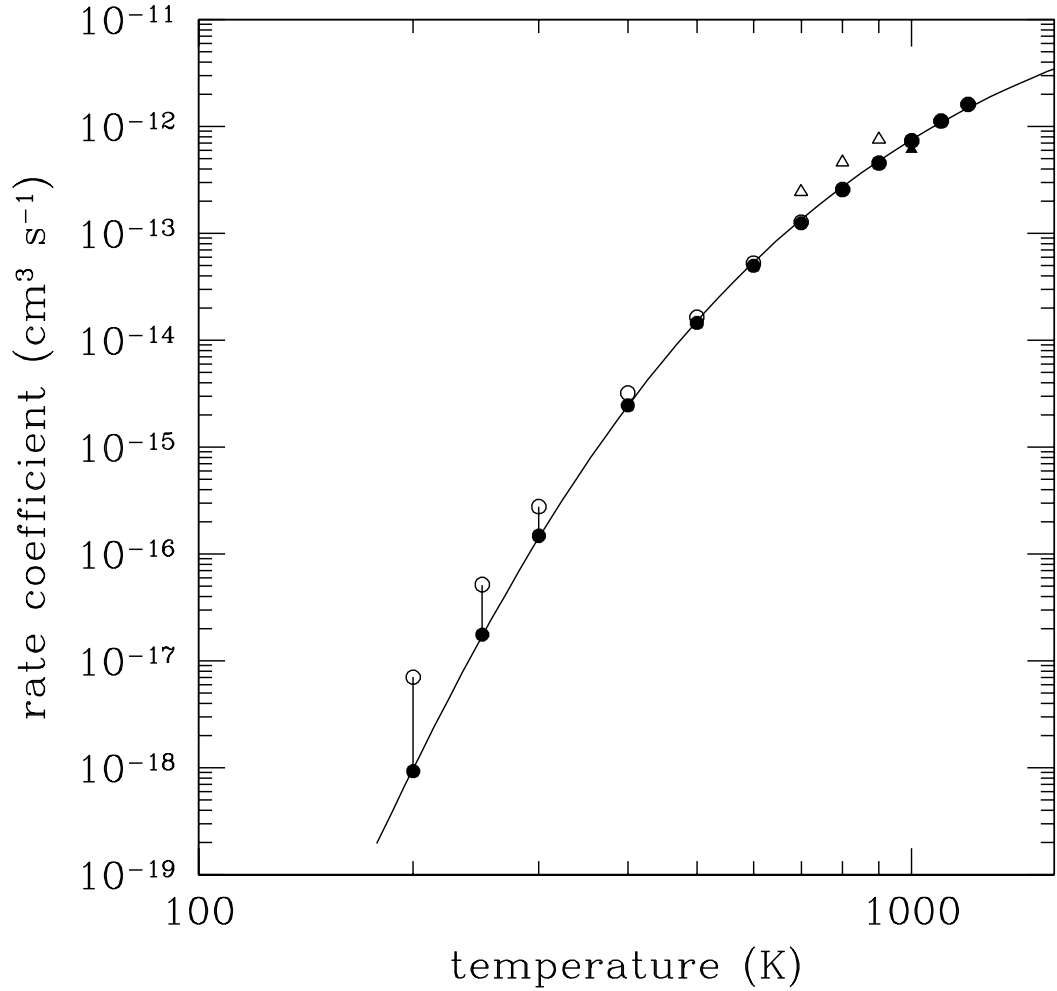}
\caption{Rate coefficient for the reaction $\rm{HD}+{\rm H} \rightarrow
{\rm D}+{\rm H}_2$ according to the calculations of Shavitt~(1959) for
two values of the asymmetric stretching force constant $A_u$ ({\it
filled dots}: $A_u=-0.358\times 10^5$~dyne~cm$^{-1}$; {\it empty dots}:
$A_u=-0.732\times 10^5$~dyne~cm$^{-1}$). The experimental results of
Boato et al.~(1956) and van Meersche~(1951) are shown by {\it filled
triangles} and {\it empty triangles}, respectively. The {\it solid
line} is our fit to the data of Shavitt~(1959) for $A_u=-0.358\times
10^5$~dyne/cm$^{-1}$ (see Table~1). }
\end{figure}
\clearpage

\begin{figure}
\plotone{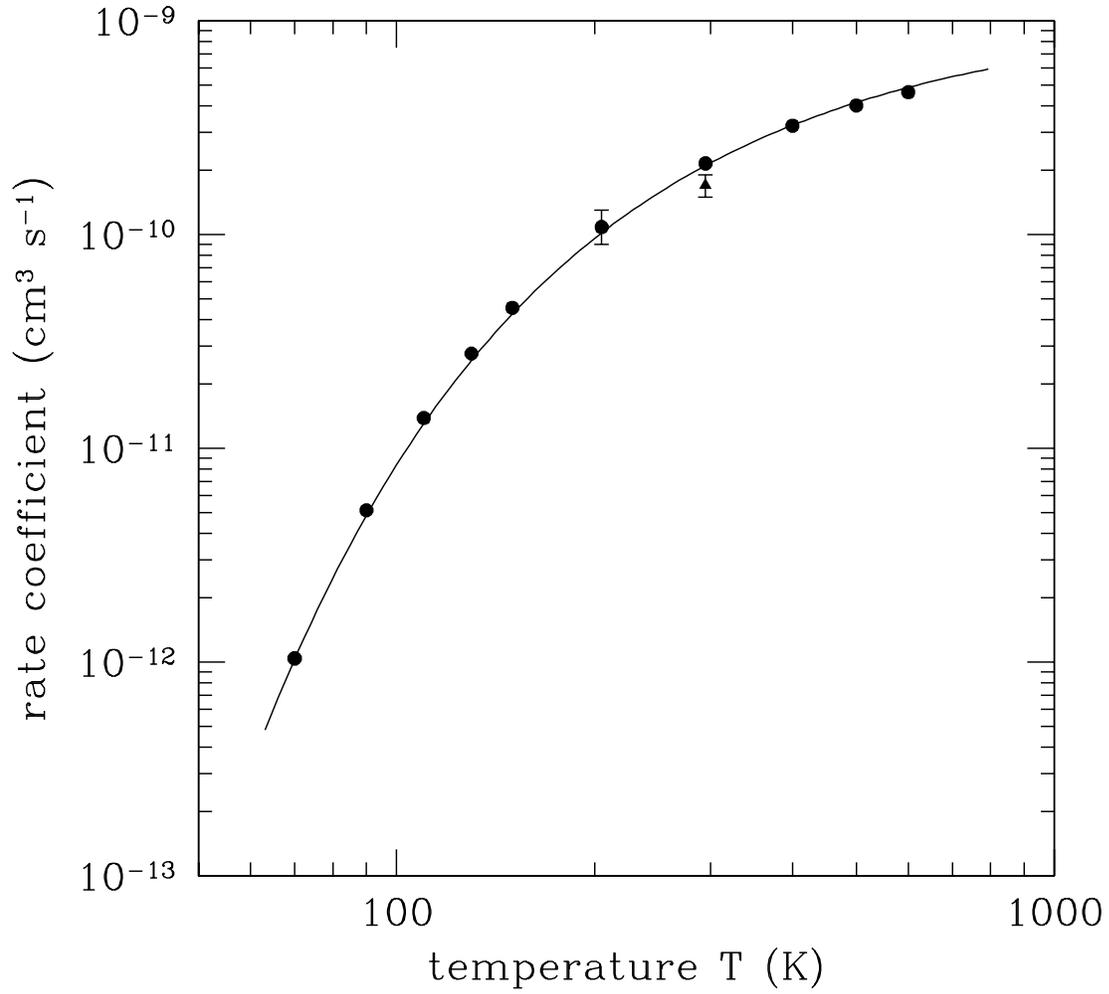}
\caption{Rate coefficient for the reaction $\rm{HD}+{\rm H}^+
\rightarrow {\rm D}^+ +{\rm H}_2$ according to the calculations of
Gerlich~(1982) ({\it filled dots} and the exprimental results of
Henchman et al.~(1981) at $T=205$ and 295~K ({\it triangles with
errorbars}). The {\it solid line} shows our fit to Gerlich's results 
(see Table~1).}
\end{figure}
\clearpage

\begin{figure}
\plotone{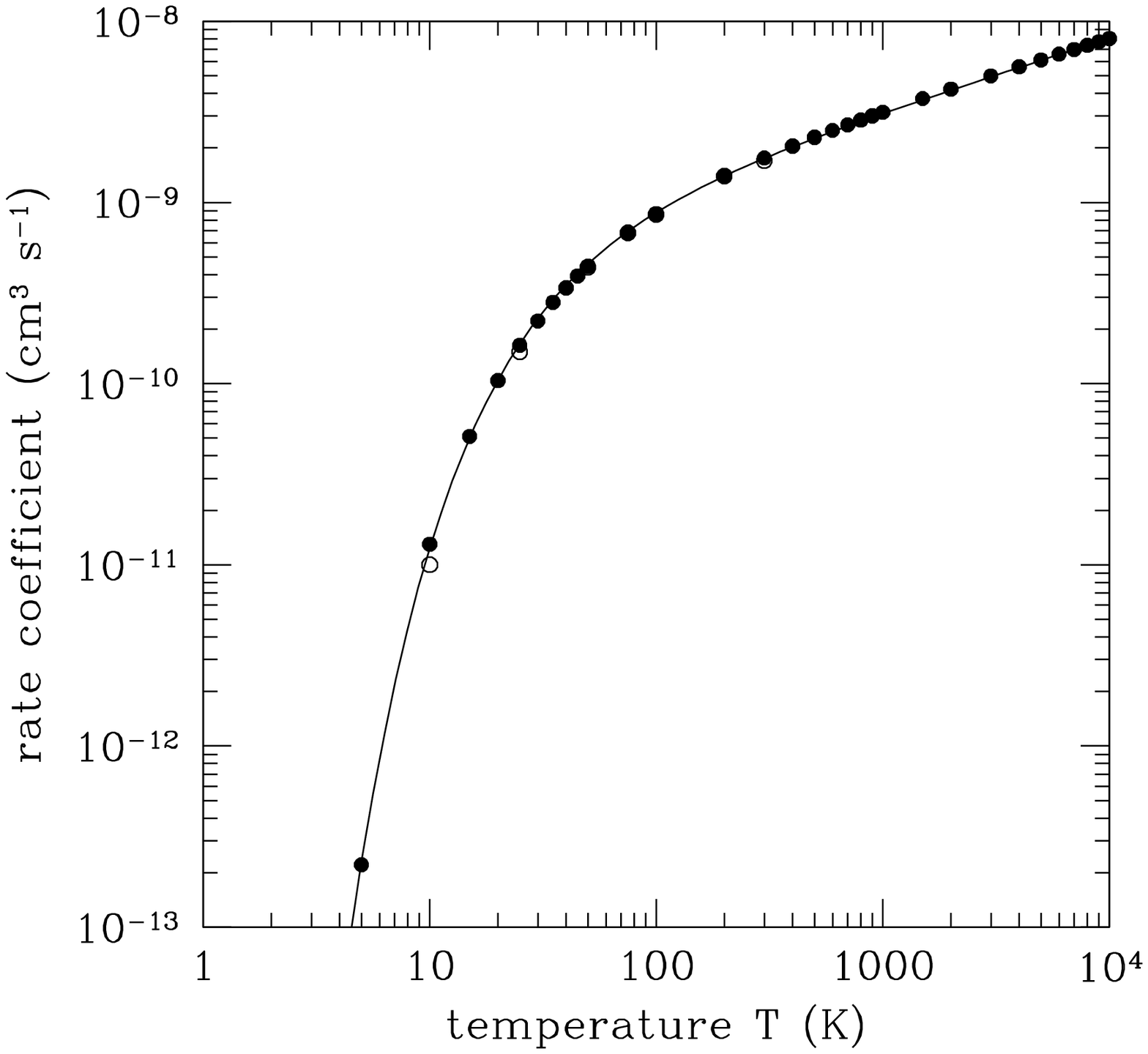}
\caption{Rate coefficient for the reaction ${\rm H}^++{\rm D}
\rightarrow {\rm H}+{\rm D}^+$ according to the calculations of Wolf
Savin~(2001) ({\it filled dots}) and Watson et al.~(1978) ({\it empty
dots}). The {\it solid line} shows the fit obtained by Wolf Savin~(2001) 
to his results (see Table~1).} 
\end{figure}
\clearpage

\begin{figure}
\plotone{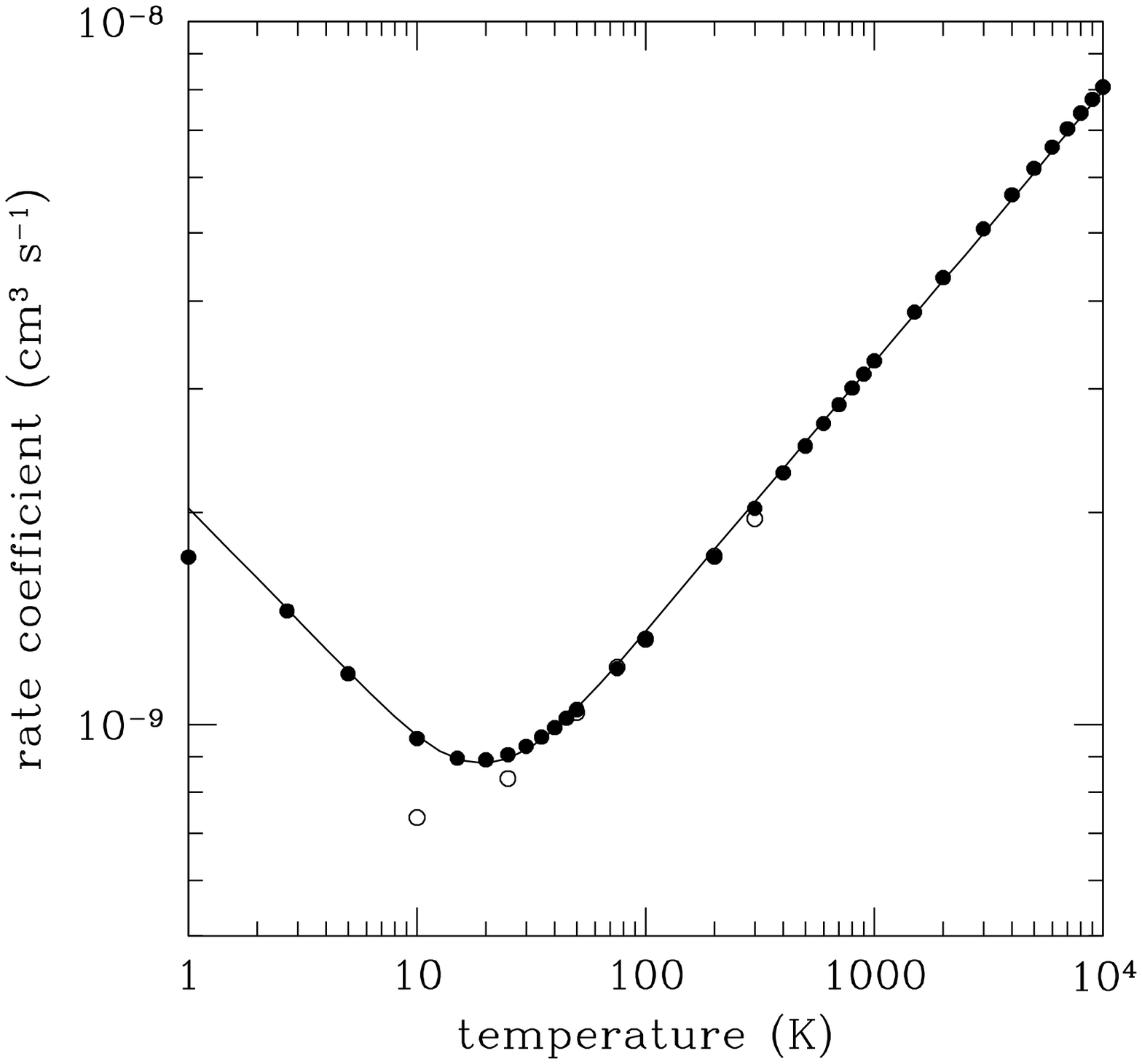} \caption{Rate coefficient for the reaction ${\rm
D}^++{\rm H} \rightarrow {\rm D}+{\rm H}^+$ according to the
calculations of Wolf Savin~(2001) ({\it filled dots}) and Watson et
al.~(1978) ({\it empty dots}). The {\it solid line} shows the fit obtained by
Wolf Savin~(2001) to his results (see Table~1).}
\end{figure}
\clearpage

\begin{figure}
\plotone{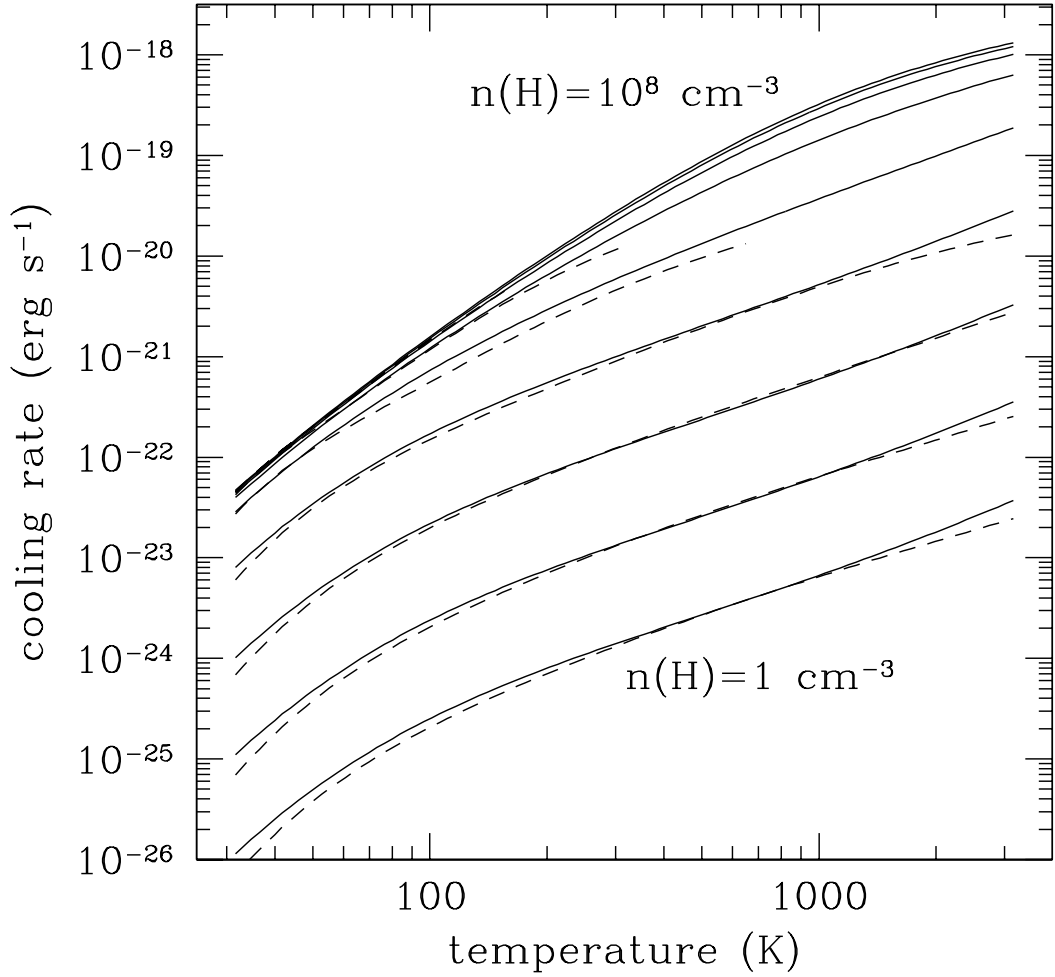} \caption{The cooling rate per HD molecule
computed for $n({\rm H})=1$ to $10^8$~cm$^{-3}$ ({\it solid lines}:
Flower et al.~2000; {\it dashed lines}: GP). Only collisions of HD
with H have been considered.}
\end{figure}
\clearpage

\begin{figure}
\plotone{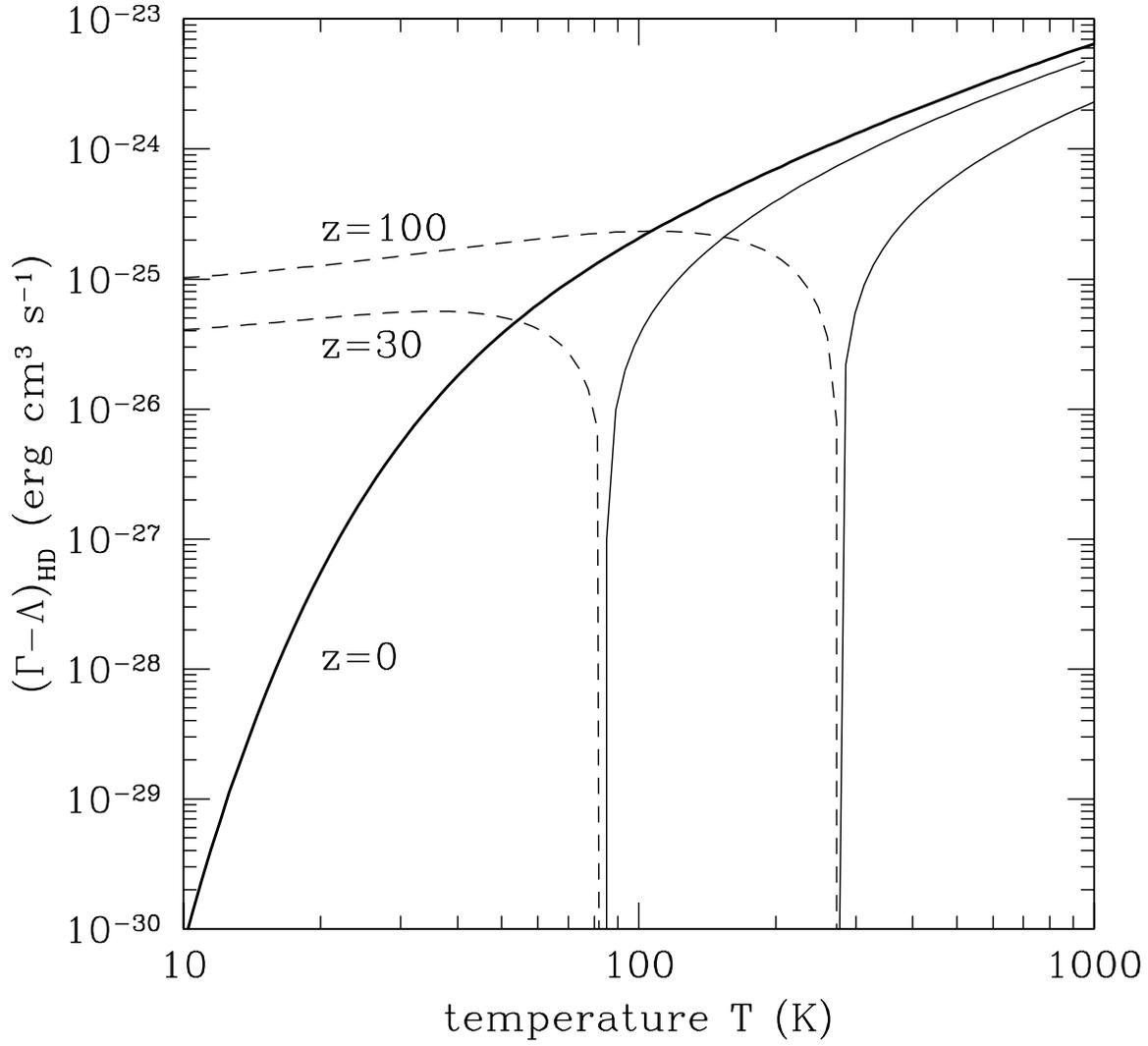}
\caption{The heat transfer function $(\Gamma-\Lambda)_{\rm HD}$ for
$n({\rm H})=1$~cm$^{-3}$ versus gas temperature at selected redshifts.
The solid line is computed ignoring the effects of the CBR ($z=0$).
When $T_{\rm gas}>T_{\rm rad}$ the heat exchange is a cooling term
({\it solid lines}).  In the opposite case, the transfer becomes a
heating source for the gas ({\it dashed lines}).}
\end{figure}
\clearpage

\begin{figure}
\plotone{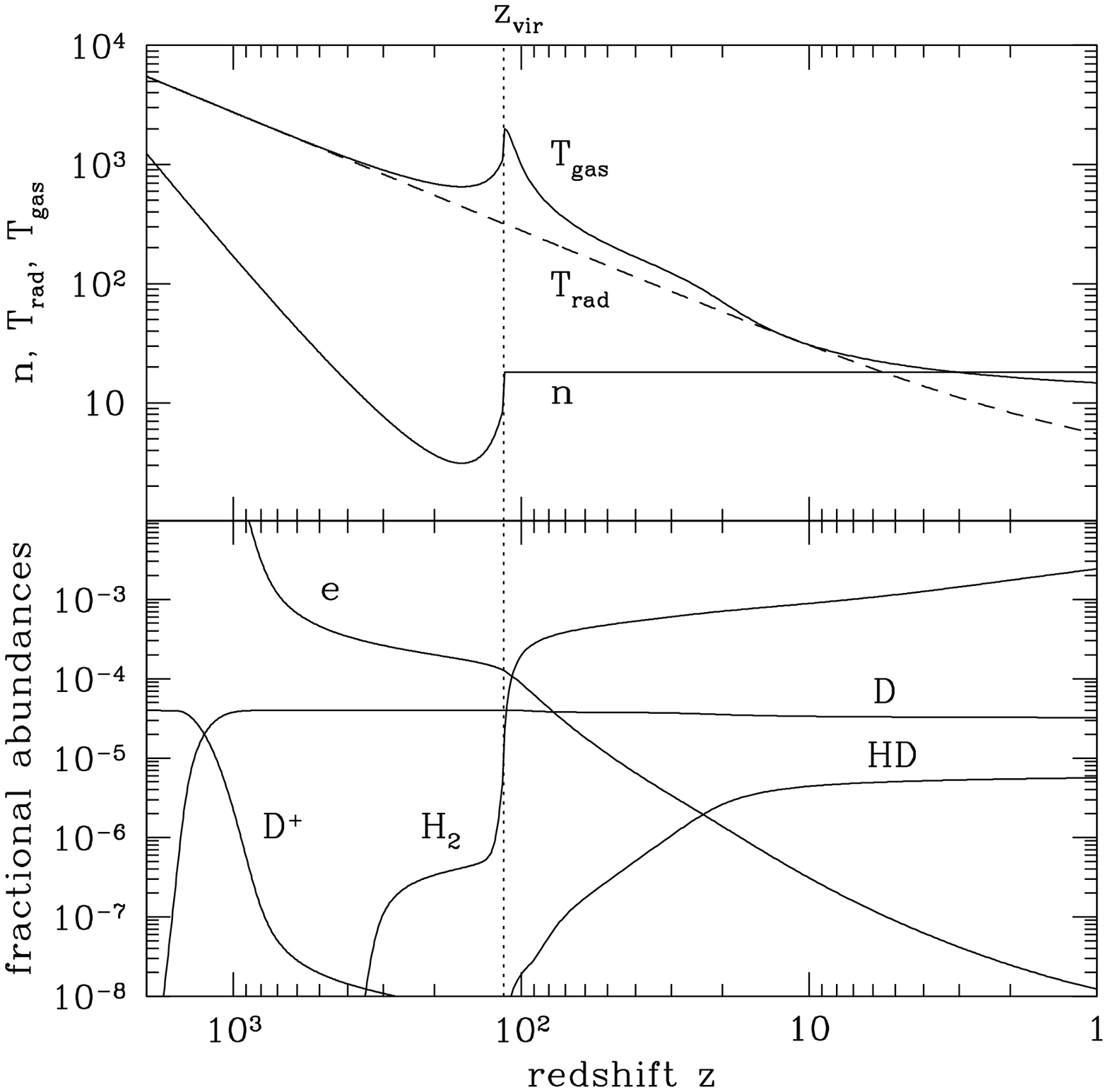}
\caption{The time evolution of gas in a primordial cloud for $z_{\rm
vir}=110$, $T_{\rm vir}=2000$~K, $h=0.5$, and $\Omega_b=0.06$. The mass 
of the cloud is $1.35\times 10^6$~$M_\odot$. The top
panel shows the gas density $n$ (in cm$^{-3}$), the gas temperature $T_{\rm gas}$ (in K)
and the temperature of the cosmic background radiation $T_{\rm rad}$ (in K) as
function of redshift $z$.  The bottom panel shows the abundances of
electrons, D, D$^+$, H$_2$ and HD, relative to H, for the same model.}
\end{figure}
\clearpage


\begin{thebibliography}{}

\bibitem[]{} Abel, T., Anninos, P., Norman, M. 1997, New Astr., 2 181

\bibitem[]{} Abgrall, H., Roueff, E., \& Viala, Y. 1982, A\&AS, 50, 505

\bibitem[]{} Berry, R. S., Rice, S. A., \& Ross, J., 1980, Physical Chemistry (New York: Wiley), p. 756

\bibitem[]{} Boato, A., Careri, G., Cimino, S., Molinari, S., \& Volpi, C. 1956, J. Chem. Phys., 24, 783

\bibitem[]{} Bromm, V., Coppi, P. S., \& Larson, R. B. 1999, ApJ, 527, L5

\bibitem[]{} Burles, S., \& Tytler, D. 1998a, ApJ, 499, 699

\bibitem[]{} Burles, S., \& Tytler, D. 1998b, ApJ, 507, 732

\bibitem[]{} Cen, R., Ostriker, J. P., \& Peebles, P. J. E. 1993, ApJ, 415, 423

\bibitem[]{} Charutz, D. M., Last, I., \& Baer, M. 1997, J. Chem. Phys. 106, 7654

\bibitem[]{} Dalgarno, A., Weisheit, J. C., \& Black, J. H. 1973, Astrophys. Lett., 14, 77

\bibitem[]{} D'Odorico, S., Dessauges-Zavadsky, M., \& Molaro, P. 2001, A\&A, 368, L21

\bibitem[]{} Esry, B. D., Sadeghpour, H. R., Wells, E., \& Ben-Itzhak, I. 2000, J. Phys. B, 33, 5329

\bibitem[]{} Fehsenfeld, F. C., Dunkin, D. B., Ferguson, E. E., \& Albritton, D. L. 1973,
             ApJ, 183, L25

\bibitem[]{} Fehsenfeld, F. C., Albritton, D. L., Bush, Y. A., Fournier, P. G., 
             Govers, T. R., \& Fournier, J. 1974, J. Chem. Phys., 61, 2150

\bibitem[]{} Flower, D. R., \& Roueff, E. 1999, MNRAS, 309, 833

\bibitem[]{} Flower, D. R. 2000, MNRAS, 318, 875

\bibitem[]{} Flower, D. R., Le Bourlot, J., Pineau des For\^ets, G., \& Roueff, E. 2000,
	     MNRAS, 314, 753

\bibitem[]{} Flower, D. R., \& Pineau des F\^orets, G. 2001, MNRAS, 323, 627

\bibitem[]{} Galli, D., \& Palla, F. 1998, A\&A, 456, 631 (GP)

\bibitem[]{} Gerlich, D., 1982, in Symp. on Atomic and Surface Physics, W. Lindinger,
             F. Howorka, T. D. M\"ark, F. Egger eds. (Dordrecht: Kluwer), p. 304

\bibitem[]{} Green, S. 1974, Physica, 76, 609

\bibitem[]{} Haiman, Z., Thoul, A. A., \& Loeb, A. 1996, ApJ, 464, 523

\bibitem[]{} Henchman, M. J., Adams, N. G., \& Smith, D. 1981, J. Chem. Phys., 75, 1201

\bibitem[]{} Hodges, R. R., Jr., \& Breig, E. L. 1993, J. Geophys. Res., 98, 1581

\bibitem[]{} Hogan, C. J. 1998, in {\it Primordial Nuclei and their Galactic evolution}, 
	     N. Prantzos, M. Tosi, \& R. von Steiger eds. (Dordrecht: Kluwer), p. 127

\bibitem[]{} Hunter, G., \& Kuriyan, M. 1977, Proc. R. Soc. London Ser. A, 358, 321

\bibitem[]{} Igarashi, A., \& Lin, C. D. 1999, Phys. Rev. Lett., 83, 4041

\bibitem[]{} Khersonskii, V. K. 1986, Astrophysics, 24, 114

\bibitem[]{} Lepp, S. \& Shull, J. M. 1983, ApJ, 270, 578

\bibitem[]{} Lepp, S. \& Shull, J. M. 1984, ApJ, 280, 465

\bibitem[]{} Levshakov, S. A., \& Varshalovich, D. A. 1985, MNRAS, 212, 517

\bibitem[]{} Levshakov, S. A., Molaro, P., Centuri\'on, M., D'Odorico, S., Bonifacio, P.,
	     \& Vladilo, G. 2000, A\&A, 361, 803

\bibitem[]{} Matveenko, A. V. 1974, Sov. Phys. JETP, 38, 1082

\bibitem[]{} Michael, J. V., \& Fisher, J. R. 1990, J. Phys. Chem., 93, 3318

\bibitem[]{} Michael, J. V., \& Fisher, J. R., \& Bowman, Q. S. 1990, Science, 249, 269

\bibitem[]{} Mielke, S. L., Lynch, G. C., Thrular, D. G., \& Schwenke, D. W. 1994, J. Phys. Chem., 98, 8000

\bibitem[]{} Mitchell, D. N., \& LeRoy, D. J. 1973, J. Chem. Phys., 58, 3449

\bibitem[]{} Newman, J. H., Cogan, J. D., Ziegler, D. L., Nitz, D. E., Rundel, R. D., 
	     Smith, K. A., \& Stebbings, R. F. 1982, Phys. Rev. A, 25, 2976

\bibitem[]{} O'Meara, J. M., Tytler, D., Kirkman, D., Suzuki, N., Prochaska, J. X., 
	     Lubin, D., \& Wolfe, A. M. 2001, ApJ, 522, 718

\bibitem[]{} Padmanabhan, T. 1993, {\it Structure formation in the Universe}, Cambridge Univ. Press

\bibitem[]{} Palla, F., Galli, D., \& Silk, J. 1995, ApJ, 451, 44

\bibitem[]{} Phillips, D. L., Levene, H. B., \& Valentini, J. J. 1989, J. Chem. Phys. 90, 1600

\bibitem[]{} Puy, D., Alecian, G., Le Bourlot, J., L\'eorat, J., \& Pineau des F\^orets, G.
	     1993, A\&A, 267, 337

\bibitem[]{} Ridley, B.A., Schulz, W.R., LeRoy, D.J. 1966, J. Chem. Phys., 44, 3344 

\bibitem[]{} Roueff, E., \& Zeippen, C. J. 1999, A\&A, 343, 1005

\bibitem[]{} Rozhenstein, V. B., Gershenzon, Yu. M., Ivanov, A. V., Il'in, S. D., 
             Kurcheryavii, S. I., \& Umanskii, S. Ya. 1985, Chem. Phys. Lett., 121, 89

\bibitem[]{} Schaefer, J. 1990, A\&AS, 85, 1101

\bibitem[]{} Schofield, K. 1967, Planet. Space Sci., 15, 643

\bibitem[]{} Shavitt, I. 1959, J. Chem. Phys., 31, 1359

\bibitem[]{} Smith, D., Adams, N. G., \& Alge, E. 1982, ApJ, 263, 123

\bibitem[]{} Stancil, P. C., \& Dalgarno, A. 1997, ApJ, 490, 76

\bibitem[]{} Stancil, P. C., Lepp, S., \& Dalgarno, A. 1998, ApJ, 509, 1 (SLD)

\bibitem[]{} Tegmark, M., Silk, J., Rees, M. J., Blanchard, A., Abel, T., \& Palla, F. 1997, ApJ, 474, 1

\bibitem[]{} Timmermann, R. 1996, ApJ, 456, 631

\bibitem[]{} Uehara, H., \& Inutsuka, S. 2000, ApJ, 531, L91

\bibitem[]{} van Meersche, M. 1951, Bull. Soc. Chim. Belges, 60, 99

\bibitem[]{} Varandas, A. J. C., Brown, F. B., Mead, C. A., Thrular, D. G., \& Blais, N. C. 1987, 
	     J. Chem. Phys., 86, 6258

\bibitem[]{} Watson, W. D. 1976, Rev. Mod. Phys., 48, 513

\bibitem[]{} Watson, W. D., Christensen, R. B., \& Deissler, R. J. 1978, A\&A, 65, 159

\bibitem[]{} Westenberg, A. A., \& de Haas, N. 1967, J. Chem. Phys., 47, 1393

\bibitem[]{} Wolf Savin, D. 2001, ApJ, in press (astro-ph/0109356)

\bibitem[]{} Zhang, J. Z. H., \& Miller, W. H. 1989, J. Chem. Phys., 91, 1528

\bibitem[]{} Zhao, Z. X., Igarashi, A., \& Lin, C. D. 2000, Phys. Rev. A, 62, 042706

\end{thebibliography}
\end{document}